# Developing Cyber Buffer Zones

Michael Robinson[1], Kevin Jones[1], Helge Janicke[2] and Leandros Maglaras[2]

[1]Airbus Group, U.K.

[2]DE Montfort University, U.K.

## INTRODUCTION

Cyberspace has become the latest domain of war (Robinson, Jones, & Janicke, 2015), where modern international actors aggressively pursue their national security and foreign policy goals (Martins, 2018). Much research has been focused upon this area, covering topics such as the ethics of cyberwarfare (Dipert, 2010), legal aspects (Baradaran & Habibi, 2017) and how best to conduct military operations inside of cyberspace (Liles, Rogers, Dietz, & Larson, 2012). Surveys of the literature show a vast range of additional topics, demonstrating that research interest into cyber warfare is lively and diverse (Robinson, Jones, & Janicke, 2015). Whilst it is clear that interest in cyber warfare is high, there has been less attention paid to its aftermath. What effects on societies persist after cyber warfare and do these effects stymie work to restore peace and security to conflict torn regions?

The field of cyber peacekeeping addresses these questions, looking at conflicts which contain cyber warfare through the lens of peacekeeping. In this chapter, we provide a background to cyber peacekeeping and survey existing literature. We then make a contribution to the field by developing the concept of a cyber buffer zone.

## BACKGROUND

The concept of cyber peacekeeping can be traced back to an article by Cahill, Rozinov and Mule (2003). They noted that cyber warfare would likely have devastating effects well beyond the boundaries of the combatants and that some kind of peacekeeping capability in cyberspace would be needed (Cahill, Rozinov, & Mule, 2003). Some potential cyber peacekeeping activities were proposed, such as cyber border management and monitoring/verification and their overall approach was to explore how existing peacekeeping doctrine could be mapped to cyber warfare. The topic did not receive any further attention for ten years until Kleffner and Dinniss (2013) reignited discussion. They drew attention to the convergence of two significant global trends: an increase in conflicts which involve a cyber component and the increasing deployment of complex peace operations. They noted that these trends made it natural to assume that peacekeepers will find themselves asked to keep the peace in environments where the peace is threatened by cyber incidents (Kleffner & Dinniss, 2013).

Akatyev and James (2015) contributed by proposing a cyber peacekeeping model, including a set of goals and proposals of activities to perform during three stages: no conflict, conflict and post-conflict. In this regard, the model covers all three phases of warfare with the primary goal of protecting civilians. The need for cyber peacekeeping was reinforced two years later by Dorn (2017), who states that cyberpeackeepers could patrol and act in cyberspace just as current UN peacekeepers patrol and act in the world's conflict zones. Faced with a huge disaster bill and a potential for vast escalation in attacks, an investment in cyber peacekeeping would seem like a bargain (Dorn, 2017).

In 2018, Robinson et al. (2018) built upon the foundations set by Cahill, Rozinov and Mule back in 2003. They reinforced the need for cyber peacekeeping with specific cases where cyber warfare would present a threat to international peace and security as defined by the United Nations. They explored how the activities of a modern multi-dimensional peacekeeping operation could be translated into a cyber warfare context, and evaluated each one according to two core criteria: value and feasibility. Any activity performed during cyber peacekeeping must bring clear value towards restoring peace and security, and must also be feasible to perform. They conclude that many of the existing UN peacekeeping activities would bring value in a cyber warfare context, but that feasibility can vary due to technical and political constraints.

## FOCUS OF THE ARTICLE

In this article, we build upon the work of Robinson et al. (2018) by taking a closer examination of just one of the proposed activities: cyber buffer zones. We begin with a brief background of traditional physical buffer zones as used by UN peacekeeping, before progressing to propose how a cyber buffer zone would be constructed and how it could bring value.

Interposition as a buffer zone is one of the core activities of a traditional UN peacekeeping operation. The UN defines a buffer zone as "an area established between belligerents and civilians that is protected and monitored by battalion peacekeeping forces and where disputing or belligerent forces and attacks on each other and the civilian population have been excluded" (United Nations, 2012). Peacekeepers keep this buffer zone free from military personnel, weapons, installations and activities and have the authority to use force in order to protect the safety and integrity of the buffer zone (United Nations, 2012).

Robinson et al. (2018) adapted this definition to produce a definition of a cyber buffer zone: "A network or site that is protected and monitored by peacekeeping forces, where cyber attacks have been excluded." They proposed that a cyber buffer zone would be most valuable at sites where the greatest threats to peace exist, and an example of critical national infrastructure was given. These are sites that civilians rely upon for essential services such as electricity, water and transport. A cyber buffer zone could be deployed reactively (i.e. after a suspected attack) or proactively (preventive deployment in anticipation of an attack). The literature regarding buffer zones was distilled down to two core pieces of value which need to be preserved when designing cyber buffer zones:

1. Deterrence: potential attackers are deterred from initiating an attack, due to the high probability of being detected and intercepted (increased chance of failure).
2. Robust force projection: When deterrence fails, the consequences for an attacker are significant e.g. incarceration, injury or death (increased consequence of failure).

It was suggested that deterrence in a cyber buffer zone can feasibly be achieved in a number of ways. Raising awareness that a particular site is under peacekeeping protection, knowledge that the site will be monitored by highly capable staff, and the increased likelihood of tracing attacks back to their origin could all have the potential to act as a deterrent.

With regards to projecting robust force, an approach which uses both short and long term activities was recommended. Further development of these activities was cited as future work, and this is where this chapter places its focus: proposing how such a cyber buffer zone would technically operate in order to achieve robust force projection and ultimately contribute towards maintaining peace.

## SOLUTIONS AND RECOMMENDATIONS

To begin, it is prudent to first become familiar with the activities of a traditional physical buffer zone. The UN has been operating buffer zones for many decades, and this experience provides us with a solid foundation from which to consider cyber buffer zones. The activities they perform in a physical buffer zone have been refined and improved over the years, based upon field experience. We therefore can assume that any activity performed at a UN buffer zone is proven to have value towards restoring peace and is feasible in at least some cases.

The UN Infantry Battalion Manual Volume II (United Nations, 2012) provides a list of the specific tasks that peacekeepers running a buffer zone perform, along with a description of what each task involves. These are presented in Table 1.

| TASK | DESCRIPTION |
|---|---|
| **Tactical Deployment** | Deploy tactical sub-units and detachments (both permanent and temporary) to effectively cover the entire frontage. |
| **Monitoring** | Observe, monitor, supervise and verify the cessation of hostilities/ceasefire/Truce/armistice agreements, compliance of agreements, troop deployments, etc. |
| **Interposition** | Interpose between opposing forces to stabilize the situation, where formal peace agreements are not in force. |
| **Supervision** | Supervise the implementation of the disengagement agreement. |
| **Repositioning of Belligerent Forces** | Accompany and support opposing forces to redeploy/withdraw to agreed dispositions and subsequent adherence to military status quo. |
| **Control of BZ** | Ensure no presence of military personnel, weapons, installations and activities, assist in securing the respective areas/line to prevent/intervene entry/intrusion without consent of military personnel, arms or related material in the Buffer Zone. |
| **Civilian Activities** | Monitor Crossing/Control Points across buffer zone for safe and orderly passage through by civilians in conjunction with opposing forces. Facilitate daily subsistence and routine activities of civilians in the buffer zone. |
| **Contain** | Prevent/contain violations/cross border attacks/isolated incidents taking place and if taken place, prevent it from escalating in to major conflicts. |
| **Investigations** | Follow up on complaints by investigations |
| **Proactive Deployment** | Proactive troop deployment to prevent an incident or its recurrence |
| **Area of Limitations** | Visit, monitor and ascertain compliance of activities periodically in stipulated "Areas of Limitations" (where military restrictions on deployment of body of troops and weapons systems and massing of troops not permitted). |
| **Interface and Coordination** | Act as go-betweens for the hostile parties with good liaison, close contact and effective coordination. |
| **Assist Establishment of Local Authority** | Assist/coordinate with local Government/belligerent parties in restoring its effective authority in respective areas |
| **Assist in Good** | Facilitate good governance in the area of separation/buffer zone, contribute to maintenance and restoration of |
| **Governance** | law and order and policing, establish interface with the inhabitants and help resumption of routine civilian activity (farming, electricity, water, medical support) for establishing normalcy. |
| **Assist other entities** | Assist/support formed military police elements, formed UN police elements/UNPOL, UN agencies in the area and other international organizations when tasked. |
| **Mine Awareness** | Support mine awareness, identify and mark minefields, and help in clearance of mines and unexploded ordnance. |
| **Facilitate Humanitarian Access** | Extend assistance to help ensure humanitarian access to civilian populations, provision of medical aid and facilitate voluntary and safe return of displaced personnel |
| **Reconciliation and Rapprochement** | Play an active and constructive role which is critical in preventing a recurrence of hostilities/to prevent flash point, detrimental to the peace process and work towards a comprehensive political solution. |
| **Assist Negotiation and Mediation** | Assist UN mediator and undertake mediation and negotiation when tasked or required. |
| **Other Activities** | Facilitate exchange of prisoners, refugees, IDPs, dead bodies and to retrieve livestock. |

**Table 1: UN buffer zone activities**

Looking at Table 1 we can see that the value a buffer zone brings goes beyond just deterrence and robust force projection. It in fact serves to support multiple other peacekeeping activities ranging from humanitarian assistance to peacemaking. This emphasises the view that the activities of peacekeeping are synergistic: no one activity in itself can restore peace, but together they all work towards a common goal. When designing a cyber buffer zone, opportunities for synergy with the wider peacekeeping operation must be highlighted and encouraged.

An initial practical step in considering the design of cyber buffer zones is to determine if these existing activities could map to a cyber warfare context in a way that would be valuable towards restoring peace. We take the same table and colour each activity: green if the activity holds potential value in cyber warfare, red if not. The description box is used to show our justification.

| TASK | DESCRIPTION FOR CYBER BUFFER ZONE |
|---|---|
| **Tactical Deployment** | Deployment of cyber peacekeeping resources to effectively cover the network/site. E.g. experts, monitoring tools. |
| **Monitoring** | Use of cyber peacekeeping resources (examples above) to monitor cyber related ceasefire terms. |
| **Interposition** | Defend the network from cyber attacks: perform active cyber defence |
| **Supervision** | Dependent upon disengagement agreement. But some potential in a cyber context e.g. neutral observation of collaborative efforts to restore control of network to rightful owner or collaboratively remove malware as part of a political agreement. |
| **Repositioning of Belligerent Forces** | Cyber troops are difficult to observe and can attack from anywhere. Attempting to monitor or guide their repositioning would not be valuable in a cyber context |
| **Control of BZ** | Ensure no presence of cyber attacks and malware. Regular scanning of the infrastructure to ensure it remains clean and free of malicious activity/content. |
| **Civilian Activities** | Monitor cyber infrastructure for safe and orderly use by civilians for peaceful purposes. Some infrastructure will be used by civilians for peaceful and essential services. Peacekeepers can ensure the system remains available use e.g. banking, commerce, and government services. |
| **Contain** | Prevent or contain cyber attacks from impacting the infrastructure. Isolating and neutralising cyber attacks to avoid escalation into further conflict or relapse into warfare. |
| **Investigations** | Follow up on complaints of cyber attack/malware infections (cyber forensic capability) |
| **Proactive Deployment** | Proactive deployment of cyber defences to pre-empt and thwart potential attacks |
| **Area of Limitations** | Spot checks of networks to ensure that they are not compromised. This could relate to networks that do not have a full time cyber peacekeeping presence but should still be occasionally checked for compromise |
| **Interface and Coordination** | Cyber Peacekeepers in a cyber buffer zone can act as go-betweens for the hostile parties |
| **Assist Establishment of Local Authority** | Assisting the rightful owner of the network in regaining control e.g. restoring power grid and ensuring no unauthorised access remains |
| **Assist in Good Governance** | Advise network owner on how to best protect the network and potentially go beyond security to provide performance/optimisation advice. Long term cyber security capacity building. |
| **Assist other entities** | Assist police and other entities e.g. assisting with trace backs, providing cyber forensic capability in case of cybercrime. |
| **Mine Awareness** | Identify and contain malware, assist in malware clearance in the network and report on possible malware outside of infrastructure (e.g. detection of external botnet for other teams to follow up on) |
| **Facilitate Humanitarian Access** | Potential to protect hospitals from cyber attack, keep routes open (e.g. airport systems, air traffic control) and ensure basic services such as power and water to help humanitarian efforts succeed. |
| **Reconciliation and Rapprochement** | Cyber peacekeepers in a cyber buffer zone are unlikely to perform reconciliation duties |
| **Assist Negotiation and Mediation** | Some potential to assist wider negotiations through detailed knowledge of the network. E.g. How exposed it is, current events, when can the cyber buffer zone be withdrawn etc. |
| **Other Activities** | N/A to cyber peacekeeping |

Table 2: Existing buffer zone activities mapped to cyberspace

Studying Table 2, many of the existing buffer zone activities have potential to bring value as part of a cyber buffer zone. But this is only one piece of the puzzle: it shows there is value to be found but not how to feasibly achieve that value. The second piece of the cyber buffer zone puzzle is how to implement the cyber buffer zone and realise that value in a feasible way.

Looking to the field of cyber security, the concept of securing some infrastructure in the face of cyber attacks is not new. The domain of cyber incident response is a well-established field which can serve as a foundation for implementing a cyber buffer zone. The NIST computer security incident handling guide (Paul Cichonski, 2012) sets out four main tasks of incident response:

- Preparation
- Detection and analysis
- Containment eradication and recovery
- Post incident activity

These guidelines provide a good foundation but do not necessarily apply well to critical infrastructure (Ying, Maglaras, Janicke, & Jones, 2015) or indeed to a cyber buffer zone. For example, in regards to preparation it is suggested that network diagrams and lists of critical assets be created and stored. In the context of a reactive cyber buffer zone, peacekeepers may not have time to collect and study such documents. However, other parts are useful. For example we are reminded that a forensic capability is not only important for later legal proceedings, but also in the immediate concern of providing insight into the current state of the system (Eden, et al., 2016).

The US based ICS-CERT (United States Homeland Security, 2009) provides guidance which is more tailored towards industrial control systems and much of the guidance provided can be adapted to help design cyber buffer zones. For example, it describes a number of staffing roles that are necessary including control system engineers, network specialists, system administrators, legal experts and vendor support specialists. This highlights how peacekeepers running a cyber buffer zone must bring expertise from a range of domains to protect the site. However, while we can use this guidance it is also not specific to a peacekeeping environment where one of the primary goals is the protection of civilians (UN Department of Peacekeeping Operations, 2015). We therefore propose a set of activities for a cyber buffer zone, built using existing guidance where it makes sense but with new proposals where necessary to suit the goals and conditions of peacekeeping.

Peacekeepers at a cyber buffer zone will conduct two sets of activities following its establishment:

- Phase One: Immediate Activities
- Phase Two: Longer Term Activities

**PHASE ONE**

Phase one consists of the immediate activities. These are rapid and decisive actions designed to gain situational awareness, stabilise the site, address any critical conditions which are an immediate threat to peace and security and restore essential services.

The phase begins by leveraging an existing concept from peacekeeping: a technical assessment mission (TAM) (United Nations Department of Peacekeeping Operations, 2014). In a proactive deployment, this will be a small team of experts who visit the site and assess its suitability for a cyber buffer zone. Since it is a proactive deployment, time will be available to become familiar with the site, collect network maps, asset lists, note observed problems and estimate the expertise that will be required. This is in line with existing guidance relating to the preparation phase of cyber incident response. In a reactive deployment, less time for preparation is available. A rapid response team may be a suitable way to address this problem: a team of peacekeepers who are able to quickly enter a site and perform a rapid assessment. The outcome is to gain situational awareness of the infrastructure, the resources required to secure and stabilise it and a determination on if the site would be suitable for a cyber buffer zone. Some key questions the TAM must answer include:

- Does the peace operation have the required resources to effectively run a cyber buffer zone at this site?
- Would a cyber buffer zone at this site provide value to the wider peace operation (e.g. could failure of the site threaten civilians, lead to state collapse, hamper restoration of state authority…)

If the TAM concludes that the site would be suitable for a cyber buffer zone, a team is established consisting of the required expertise. The process for establishing this team and ways of working (e.g. creation of a war room and communication procedures) can follow best practices described in the ICS-CERT guidance (United States Homeland Security, 2009).

The maintenance of peace and security must remain the overarching priority for a cyber buffer zone. We therefore propose that after establishment, a cyber buffer zone will conduct the following immediate tasks:

**Stabilisation** – In reactive deployments, a site such as a power plant or water treatment plant may currently be under attack, remotely via a network or locally via malware or insider. An essential task will be to prevent the condition of the site deteriorating. By this we mean working to prevent additional damage which could lengthen the amount of downtime or present a physical threat (e.g. explosion or contamination of water supply). Both of these are critical to avoid, since they are direct threats to peace and security. In proactive deployments where no negative effects are yet observed, the site will already be in a stable condition. Stabilisation will involve the following subtasks:

1. **Situational Awareness** – Both in reactive and proactive deployments, peacekeepers will need to quickly gain situational awareness: an understanding of the infrastructure, its assets, interdependencies, protocols and people. Information collected by the TAM will be useful here as a foundation. Confirmation of what is working and what is failing. Triage of expertise to specific components of the infrastructure for further analysis on the cause of the failures e.g. malware, exploited vulnerability, hardware fault, human error, malicious insider. In the case of attack, the team must determine the attack vector used.
2. **Action Plan** – Once the cyber buffer zone team has gained awareness of the infrastructure, an understanding of where failures are and attack vectors used an action plan can be developed. This plan will aim to directly address the most critical obstacles towards the site being stabilised and assign resources to actions. For example, vendor support experts may be assigned to replace failed hardware whilst malware and forensics specialists work to identify and neutralise malware on engineering stations. Others will be assigned to close the specific vulnerabilities used in attacks. The action plan must be discussed and agreed with all stakeholders, including local staff. Note that tasks one and two will not always be linear: following the action plan the team reassess the site and modify the plan based on new information or a change in the environment.

Phase one can be considered successful when there are no remaining immediate threats to civilians, damage is not spreading, malware has been neutralised (e.g. contained, command and control cut) and actively exploited vulnerabilities have been closed. The team can then proceed to phase two: longer term activities.

**PHASE TWO**

The goals of this phase are to restore service and establish a long term security capability at the site.

1. **Restoration of Service** – Once a site has been stabilised the next goal is to restore operation and resume provision of the service. Again this will require input from a number of areas including local staff, vendors and experts from the industry in question.

2. **Active defence** - Once service has been resumed, it will be necessary to monitor the infrastructure and defend against any new attacks. Peacekeepers will deploy cyber defences: IP address blocking, load balancing, reconfiguration and redirecting of denial of service traffic are all examples possible defences. Companies such as Google have expressed interest in assisting the mitigation of denial of service attacks for good causes such as election monitoring groups and human rights organisations

(Google, 2016). Organisations such as this would make good partners for a cyber buffer zone and demonstrates how important technology partners could be in making cyber buffer zones effective.

3. **Vulnerability assessment and hardening** – With the site stabilised, the team can begin the process of identifying other weaknesses in the security of the infrastructure: known vulnerable or unnecessary software/services, misconfigurations, weak credentials, unnecessary external connections, weak policies, security blindspots, gaps in physical security or human resource security. These are vulnerabilities which threaten the ongoing security of the site, but were not actively being exploited in phase one. Due to the fragile nature of industrial control systems (Wedgbury & Jones, 2015), vulnerabilities will not be closed until a plan is developed in collaboration with all stakeholders including local staff. The reason for this is twofold: first because patching software or changing long standing configurations may have unexpected consequences upon the operation of the site. Second, local stakeholders need to be engaged and involved in decision making to ensure ongoing cooperation and success of the buffer zone. The end result will be a hardened infrastructure, resilient to future attacks.

4. **Restoration of Control** – The immediate activities focused upon protecting civilians by tackling critical threats, not upon restoring all systems back to the control of the legitimate owner. In this regard, non-critical systems at the site might still be under attacker control even after phase one has completed. This activity aims to restore full control of all systems at the site back to the legitimate owner.

5. **Monitoring and Supervision** - With the immediate threats countered, cyber peacekeepers monitor the site and supervise collaborative efforts e.g. collaboration between two parties to remove malware, arranged as part of a peace agreement.

6. **Cyber Forensics** - Cyber peacekeepers can begin to provide forensics services to identify how breaches occurred and what information was stolen etc. This is useful for verifying ceasefire agreements and for third party enquiries such as national police.

7. **Peace Support** - Cyber peacekeepers can use their knowledge of the site and its staff to support wider peacekeeping tasks. For example they can provide evidence on compliance with cyber terms, report upon ongoing attacks from external sources, act as an intermediary between the hostile parties, support police investigations and share information on malware which may be valuable for other cyber buffer zones in a region.

8. **Training** - Local staff are trained in security best practices specific to the infrastructure they are maintaining, for example developing incident response plans relevant to industrial control systems. The goal is to build a local capability so that peacekeeping forces can proceed to phase three and withdraw from the site.

Phase two activities will continue until local staff are able to maintain the security of the infrastructure without cyber peacekeeper assistance. This goal is in line with existing peacekeeping doctrine, which strongly encourages local capacity building to enable eventual withdrawal of UN forces (United Nations, 2008).

**PHASE THREE**

**Withdrawal** – The eventual goal of peacekeepers, in all domains, is to build a local security capability and hand over responsibility for security to local staff. A cyber buffer zone will be no different: peacekeepers cannot stay at the site forever, and must eventually withdraw once a local capability has been built. This local capability should be effective enough to ensure the site is well protected in the long term. Activities in phase two are designed to contribute towards this: hardening of the site and training are both activities which aim to enable eventual withdrawal of cyber peacekeeping services.

A visual representation of the three phases, along with how each activity maps to the existing activities of a physical buffer zone is presented in Figure 1.

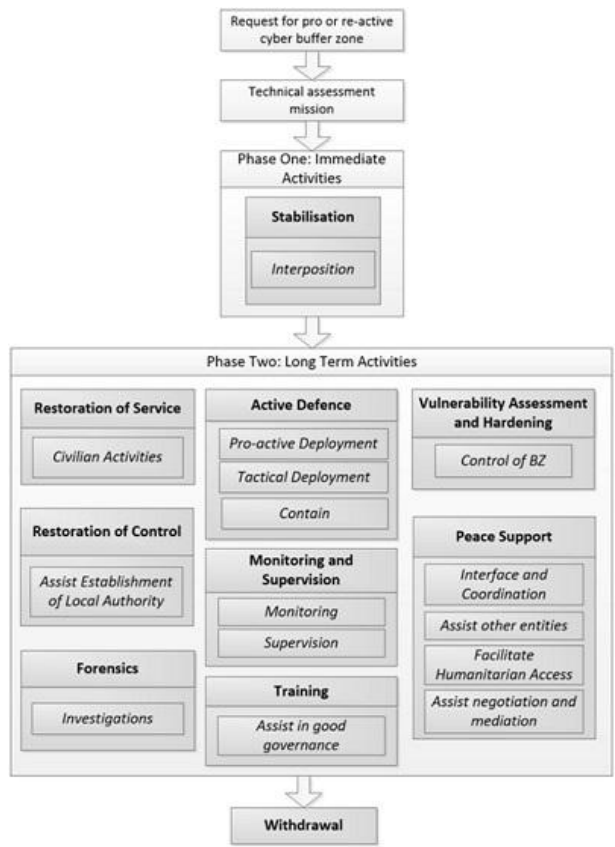

Figure 1: The three phases of a cyber buffer zone

**CONCEPT EVALUATION - FUTURE RESEARCH DIRECTIONS**

The proposals made in this article are crafted to fulfil the goals of a UN peace operation (namely, to protect civilians) whilst remaining feasible to perform. In this section, we discuss where obstacles may arise and avenues for future research.

**Local Resistance**

The success of a cyber buffer zone as described in this article is highly reliant upon the cooperation of local staff. Cyber peacekeepers will find it difficult to perform the activities we describe if local staff do not provide assistance, or actively resist. Cooperation of local staff should be measured by the technical assessment mission, and a lack of cooperation is grounds for denying a buffer zone. Unfortunately this could place the UN in a difficult situation. For example, in a case where critical infrastructure is failing to the point where it presents a threat to civilians yet local cooperation is lacking. This is a challenging scenario where future research would be welcomed.

**Lack of Consent**

While local resistance represents an unwillingness to cooperate at the field level, we must also consider a lack of consent at higher levels. Consent of the parties is one of the core principles of UN peacekeeping (United Nations, 2008), and is required for the success of the operation. Take a scenario where critical infrastructure is failing and threatening civilian security, and local staff appear to be unsuccessful in tackling the threat. The UN may feel that a cyber buffer zone is critical for peace and security in the region but may find that consent is not given by the host nation. Following existing doctrine, this naturally leads to the question of peace enforcement.

Peace enforcement has been used in the past for the following purposes (Bellamy & Williams, 2010):

- To restore or maintain international peace and security.
- To enforce sanctions.

- To defend peacekeeping personnel.
- To provide physical protection to civilians.
- To protect humanitarian activities.
- To intervene in internal conflicts.

It is therefore not radical to propose that the UN has the option to forcefully impose a cyber buffer zone at CNI which threatens the physical security of civilians or the maintenance of international peace and security. The importance of the UN's authority to enforce peace was supported by ex-secretary general Boutros-Ghali who stated that "while such action should only be taken when all peaceful means have failed, the option of taking it is essential to the credibility of the United Nations as a guarantor of international security" (United Nations, 1992). If it is agreed that the authority to enforce a cyber buffer zone exists, it must be asked if it would be feasible. This would be a valuable area for future work.

**Ensuring Impartiality**

Impartiality is the second core principle of UN peacekeeping, and must be upheld when establishing cyber buffer zones. Ensuring that a buffer zone is impartial, and is seen as such, may become a challenging task. This is because cyber peacekeepers are taking an active role: bolstering defences, training staff and restoring control. Care must be taken to ensure that any buffer zone has the clear, publicised aim of protecting civilians. It is necessary to emphasise that they will only be established at sites where a lack of protection could endanger the human rights of civilians. Military sites and systems will not be protected. It must also be made clear that a buffer zone is focused upon defence, and not offence.

A potential threat to the appearance of impartiality is that establishing a UN cyber buffer zone to protect CNI frees up that nation's cyber troops to focus upon offence rather than defence. This has the potential to lead to undesirable incentives such as letting CNI fail or purposely creating incidents so that the UN will step in and provide protection. This is a valid concern that must be considered. One response is to only provide a buffer zone once peace has been agreed. If a nation then wishes to cheat and continue offensive cyber operations while enjoying the benefits of the buffer zone, this is a risk that could result in consequences on the international stage.

The problem of resource hoarding also exists. Country A may request many cyber buffer zones simply to consume UN cyber resources and deny the service to its rival. Cyber peacekeeping resources must therefore be provisioned in a balanced manner, with restraint in mind. The technical assessment missions will be crucial here, to decline requests where there is no clear need. The protection of civilians will always be the greatest deciding factor regarding where cyber buffer zones are established.

**Securing Critical National Infrastructure**

It has been argued that cyber buffer zones will likely bring most value when established at a critical national infrastructure (CNI). It is a known issue that the hardware, software and protocols which operate CNI are particularly challenging to secure from a cyber perspective. Firstly they make attractive targets to a wide spectrum of attackers. Nation states, terrorist groups, hackers, activists, organised crime and disgruntled insiders all potentially have an interest in attacking CNI (Robinson, 2013). Secondly, the technology used at such sites is known to possess characteristics which make it particularly vulnerable to cyber attack and challenging to defend (Merabti, Kennedy, & Hurst, 2011). A discussion into all of the security issues of CNI is beyond the scope of this paper, but extensive literature exists in this area (Robinson, 2013) (Gao, et al., 2014) (Nicholson, Webber, Dyer, Patel, & Janicke, 2012). Instead, we will highlight challenges that present a specific problem to a cyber buffer zone.

Arguably one of the major challenges peacekeepers will face in building a cyber buffer zone will be to work with the hardware, software and protocols. These components can be decades old, built using proprietary technologies, fragile in their operation and designed/tested in an isolated environment. These characteristics will make establishing a cyber buffer zone at CNI much more challenging than in a traditional ICT environment.

To give a concrete example, let us consider phase one. During this phase, the peacekeepers aim to stabilise the infrastructure by identifying where critical negative effects are originating from and resolving the issue. At a traditional ICT based environment, this would involve the use of various monitoring tools. The NIST guidance suggests the use of network and host-based IDPSs, antivirus software and log analysers (Paul Cichonski, 2012). Research suggests that using such tools at CNI could have negative effects on the functioning of core components, causing them to become unresponsive, slow or unpredictable in behaviour (Coffey, Smith, Maglaras, & Janicke, 2018). It is also rare for such components to hold any logs or other forensic artefacts of value which can be retrieved (Eden, et al., 2016). Even the act of closing a vulnerability through patching firmware, software or making a configuration change could be enough to cause unexpected results which worsens the situation. The conclusion we must reach is that peacekeepers entering CNI to establish a buffer zone will require very specific skillsets in order to correctly diagnose and remediate cyber incidents without making the incident worse.

**Gaining Situational Awareness**

In the case of a reactive cyber buffer zone, peacekeepers may be asked to deploy with no or limited pre-existing knowledge of the site's hardware, software and protocols. This makes gaining situational awareness of the infrastructure an important task but difficult task. The age of the components at the site make it possible that the vendor who produced them is no longer in business. This immediately presents an obstacle towards finding expertise, documentation and replacement parts. Network diagrams may be missing or inaccurate. In the context of a warfare environment, local staff who could share knowledge may be absent. A site suffering negative effects which are directly harming civilians (e.g. faulty water treatment) will also place pressure upon the team to act quickly. Work which explores how a cyber buffer zone team can quickly gain knowledge and situational awareness of critical infrastructure under these conditions would therefore be valuable.

**Difficulty to train**

Peacekeepers working on other activities can perform training to prepare themselves for deployment in the field. For example, exercises can be run to simulate a physical buffer zone and the types of scenarios they could face. Simulating a cyber buffer zone for training purposes will be more challenging. Due to the problem of fragility at CNI and a 100% uptime requirement, the UN will have difficulty finding a site where cyber peacekeepers can test their tools and processes. Simulations are a possible solution here, but producing such a simulation, which could accurately model cyber warfare's effects on critical national infrastructure and its components is a complex task. Efforts in this area do exist (Ferreira, Machado, Costa, & Rezende, 2015) but further research would bring value to developing effective cyber buffer zones.

**Securing cyber expertise**

Cyber security expertise is in high demand across the world, with some estimating that by 2021 there will be 3.5 million unfilled cyber security roles across the public and private sector (Morgan, 2017). With threats in cyber space only rising, the demands for staff with cyber expertise will only continue to grow. This presents a challenge to the concept of cyber peacekeeping: where is the expertise going to come from and how will it be funded? Peacekeeping organisations such as the UN rely upon contributions of troops from member states, but it is questionable if states already short on cyber expertise would be willing to lose them to support a peace operation. We have also proposed that a cyber buffer will need a range of skillsets, including some which come from the private sector. It must therefore be asked what incentives these private organisations would need to contribute their expertise. Research on why states contribute towards peace operations is well established (Bove & Elia, 2011) (Bellamy & Williams, 2013), and it would be useful to explore how this body of work can be applied to the concept of cyber buffer zones.

**Political, legal and social concerns**

The concept of peacekeeping is one which touches upon multiple domains of thought. Legal, social, political and military considerations all come together to make peacekeeping a success. A cyber buffer zone will be no different. Studies which examine the legal implications and requirements will be essential. For example, it is likely that many sites critical to the protection of civilians will be owned by private entities and not the nation state itself. This raises questions about a state's ability to consent to a buffer at sites it does not outright own. Questions of privacy and data protection may also arise in the case of peacekeepers of multiple nationalities entering infrastructure to perform monitoring of network traffic. On the political aspect, highly cyber developed nations may be wary of sending their cyber experts to work alongside those from rival nations. There is even the possibility for cases where a nation is conducting or condoning cyber attacks on one hand, whilst contributing cyber peacekeepers with the other. In this regard there is potential for the goals of a cyber buffer zone to become subverted for ulterior motives, or at least for the suspicion of such to exist between nations. Research which explores and resolves these aspects is valuable towards making cyber buffer zones a success.

**CONCLUSION**

UN Peacekeeping has been operating since 1948, working to maintain peace and protect civilians around the world. As war increasingly gains a cyber component, it is important to consider how peacekeeping must adapt to remain effective. The field of cyber peacekeeping addresses this issue, and in this chapter we have made a contribution by building upon the idea of a cyber buffer zone.

Buffer zones have been used throughout the history of UN peacekeeping to physically separate conflicting forces, protect civilians and ensure peace in a region. We have proposed that the concept of a cyber buffer zone is valuable, but that there are many challenges to overcome.

From the perspective of value, a cyber buffer zone can directly address one of the main goals of UN Peacekeeping: protection of civilians. As societies become increasingly reliant upon cyber space for the provision of services such as power, water, finance, transport and commerce, the impact that cyber warfare could have upon these services rises. Failure of a smart grid, air traffic control or the rail network could at the very least prevent a nation and its citizens from recovering to a level where peace can be maintained. In some cases, it could lead to loss of life.

We have therefore designed cyber buffer zones as an activity which consists of both immediate and longer term activities. The focus of the immediate activities is to rapidly address any pressing threats to civilians and peace. Once this has been achieved, cyber security at the site is bolstered so that it remains resilient into the future.

While a cyber buffer zone is valuable, we have shown that there are many challenges to overcome. Critical national infrastructure can be difficult to secure, especially at short notice without much prior knowledge of the hardware, software and protocols in use. It will also require specific expertise that will likely be expensive in a market where cyber expertise is in high demand. Whilst these and other challenges exist, we propose that future research will be able to overcome them.

**ADDITIONAL READINGS**

**KEY TERMS AND DEFINITIONS**

**CNI**: Critical National Infrastructure.
**DDR**: Disarmament, Demobilisation and Reintegration
**UN** United Nations